\documentclass[12pt,thmsa]{revtex4}
 \usepackage{graphicx}
\begin{document}

\title{Quantum fluctuations in ultranarrow superconducting nanowires }
\author{M. Zgirski, K.-P. Riikonen, V. Touboltsev and K.Yu. Arutyunov}
\date{\today}

\begin{abstract}
Progressive reduction of the effective diameter of a nanowire is applied to
trace evolution of the shape of superconducting transition $R(T)$ in
quasi-one-dimensional aluminum structures. In nanowires with effective
diameter $\leq$ 15 nm the $R(T)$ dependences are much wider than predicted by the model of thermally activated phase slips. The effect can be explained by quantum fluctuations of the %%@
order parameter. Negative
magnetoresistance is observed in the thinest samples. Experimental results
are in reasonable agreement with existing theoretical models. The effect
should have a universal validity indicating a breakdown of zero resistance
state in a superconductor below a certain scale.
\end{abstract}

\maketitle

\address {University of Jyv\"askyl\"a, Department of Physics, NanoScience
Centre, PB 35, 40014 Jyv\"askyl\"a, Finland}

%\begin{keywords}
% nonequilidrium superconductivity, quantum fluctuations, nanostructures
%\end{keywords}

%  ############### end of front matter###############%

\newpage

\section{\protect\bigskip Introduction}

The ability to carry a dissipationless electric current is a fundamental
attribute of a superconductor. One might naively expect that this property
is preserved at reduced dimensions. Unfortunately, inevitable fluctuations
result in momentary suppression of superconductivity leading to energy
dissipation. Being integrated in time, this effect manifests itself as
finite resistance. Of particular interest are quasi-one-dimensional (quasi-1D)
systems, with effective diameter $\sigma ^{1/2}$ smaller than the
superconducting coherence length $\xi $, where there is only one parallel
channel of supercurrent. In such objects, due to fluctuations of the order
parameter, the effective resistance is never strictly equal to zero. However, a
measurable resistance at temperatures well below the critical temperature $%
T_{c}$ can be observed only in samples with rather small diameters. Early
experiments \cite{R(T)of tin whisker} confirmed that in narrow
superconducting channels the shape of the $R(T)$ transition can be explained by
thermal fluctuations \cite{Langer}. Recent experimental \cite{Giordano} - 
\cite{Altomare} and theoretical \cite{Duan} - \cite{Pesin} discoveries claim
the existence of an additional (non thermal) mechanism of finite resistance
in 1D systems. The issue is of vital importance for the development of
superconducting nanoelectronic elements designed to carry a dissipationless
electric current.

The finite resistance of a 1D channel can be understood that for sufficiently longs
system of length $L\gg \xi $ there is always the finite probability of a
fluctuation to drive instantly a fraction of the wire of volume $\xi \sigma $
into a normal state. The energy required for this process is the corresponding
superconducting condensation energy $\Delta F\sim B_{c}^{2}\xi \sigma $,
where $B_{c}$ is the critical magnetic field. If thermal effects solely
contribute to the fluctuations, it has been shown \cite{Langer} that the
effective dc voltage $V(T,I)$ is proportional to the probability of these
events $\sim \exp (-\Delta F/k_{B}T)$, where $k_{B}$ is the Boltzman
constant: 
\begin{equation}
V(T,I)=\Omega (T,I,L)\exp \left[ -\frac{\Delta F}{k_{B}T}-\left( \frac{2}{3}%
\right) ^{1/2}\frac{I^{2}}{3\pi I_{0}I_{c}}\right] \sinh \left( \frac{I}{%
2I_{0}}\right) ,
\end{equation}%
$I_{c}$ is the temperature-dependent critical current, and $I_{0}=k_{B}T/\phi
_{0}$ with $\phi _{0}$ being the superconducting flux quantum. The current
dependence is determined by the $sinh(I/2I_{0})$ term and the term under the
exponent, responsible for the effective reduction of the potential barrier $%
\Delta F$ by the finite current $I$. An exact form of the pre-factor $%
\Omega (T,I,L)$ has been proposed \cite{Langer}. Though the function $\Omega
(T,I,L)$ also contains temperature- and current-dependent terms, it
contributes negligibly compared to the strong exponential dependence.
Formally the above process can be described as the thermally activated
transition of a superconducting system from a local potential minimum to a
neighboring one separated by $\pm 2\pi $ in phase space: so called,
thermally activated phase slip (TAPS). One can make a formal analogy with a
"classical jump" of a particle over the barrier $\Delta F$ stimulated by the
thermal energy $k_{B}T$.

An alternative mechanism has been proposed associated with tunneling through
the energy barrier $\Delta F$ in phase space \cite{Giordano}. Such tunneling
to a state of lower energy, also called \textit{quantum phase slip} (QPS),
should provide an additional (to TAPS) channel of energy dissipation in a
current carrying system. For conventional type-I superconductors estimations
show that measurable deviations from the TAPS mechanism are expected in samples
with effective diameter $\sigma ^{1/2}\sim $ 10 nm. There have been a few
experiments, claiming observation of QPS phenomenon in various
superconducting materials: In, Pb and Pb-In \cite{Giordano}, Pb, Sn and
Pb-Bi \cite{Sharifi}, MoGe \cite{Bezryadin}, Sn and Zn \cite{Tian}, Al \cite%
{Zgirski NanoLett}, \cite{Altomare}. Various models consider non-thermal
mechanisms of the finite resisitivity of a 1D superconductor \cite{Saito} - \cite%
{Pesin}. In spite of the intensive research in the field, the matter is far
from being settled. From an experimental point of view, to some extent, the
inhomogeneity of a nanowire might lead to a broad superconducting transition 
$R(T)$ which can be erroneously associated with the QPS mechanism \cite%
{Zgirski PRB limitation}. Another source of misinterpretation might come
from insufficient filtering from environmental RF noise. The resulting
parasitic overheating can shift the superconducting transition to lower
temperatures, distorting the shape of the $R(T)$ dependence.

Our motivation was to perform experiments eliminating the uncertainty
related to the uniqueness of each particular nanostructure. We were able to
trace the crossover from the TAPS mechanism to QPS in the \emph{same sample}%
\textit{\ }, successively reducing its diameter between measurements, thereby
evidencing for a solely size dependent origin of the phenomenon. The
experiments performed on multiple sets of aluminum nanowires reduced by 
ion sputtering down to sub-10 nm scales showed quantitatively similar
behavior. In all structures with effective diameter $\sigma ^{1/2}<$ 15 nm
the $R(T)$ dependences deviate from the predictions of the TAPS model \cite%
{Langer}, while the QPS scenario \cite{Zaikin} provides a reasonable fit to
the data.

\section{\protect\bigskip Experiment}

In our earlier work \cite{ion sputtering} we demonstrated that low
energy $Ar^{+}$ ion sputtering progressively and non-destructively reduces
nanostructure dimensions. The penetration depth of $Ar^{+}$ ions into an $Al$
matrix at acceleration voltages of $\sim $ 500 eV is about 1.5 nm and is
comparable to the thickness of naturally formed oxide. The accuracy of the
effective diameter $\sigma ^{1/2}$ determination from the normal state
resistance, scanning electron microscope (SEM) and scanning probe microscope (SPM)
 measurements is about $\pm $ 2 nm. Only those
samples which showed no obvious geometrical imperfections were used
for further experiments. While reducing the nanowire diameter, the
low energy ion sputtering also provides "polishing" removing surface roughness
inevitable after a lift-off process \cite{Zgirski NanoLett} (Fig. 1). In
this work the method \cite{ion sputtering}, \cite{Zgirski Nanotechnology 2008} was applied to
lithographically fabricated 99.999 \% pure $Al$ nanowires with initial
cross-section of about 100 nm x 100 nm. During measurements
the structures were immersed into a directly pumped $^{4}He$ bath with 
base temperature of about 0.95 K. Contrary to Refs. \cite{Bezryadin}, \cite{Tian}, 
\cite{Altomare}, we used a four-probe configuration, eliminating necessity to
consider the contribution of the electrodes. We obtained exact quantitative
agreement between data measured with dc and ac currents using extensive
multistage rf filtering. The results are reproducible and do not suffer from
hardware-related artifacts.
\clearpage
\begin{figure}[h]
\includegraphics [width=0.9 \textwidth] {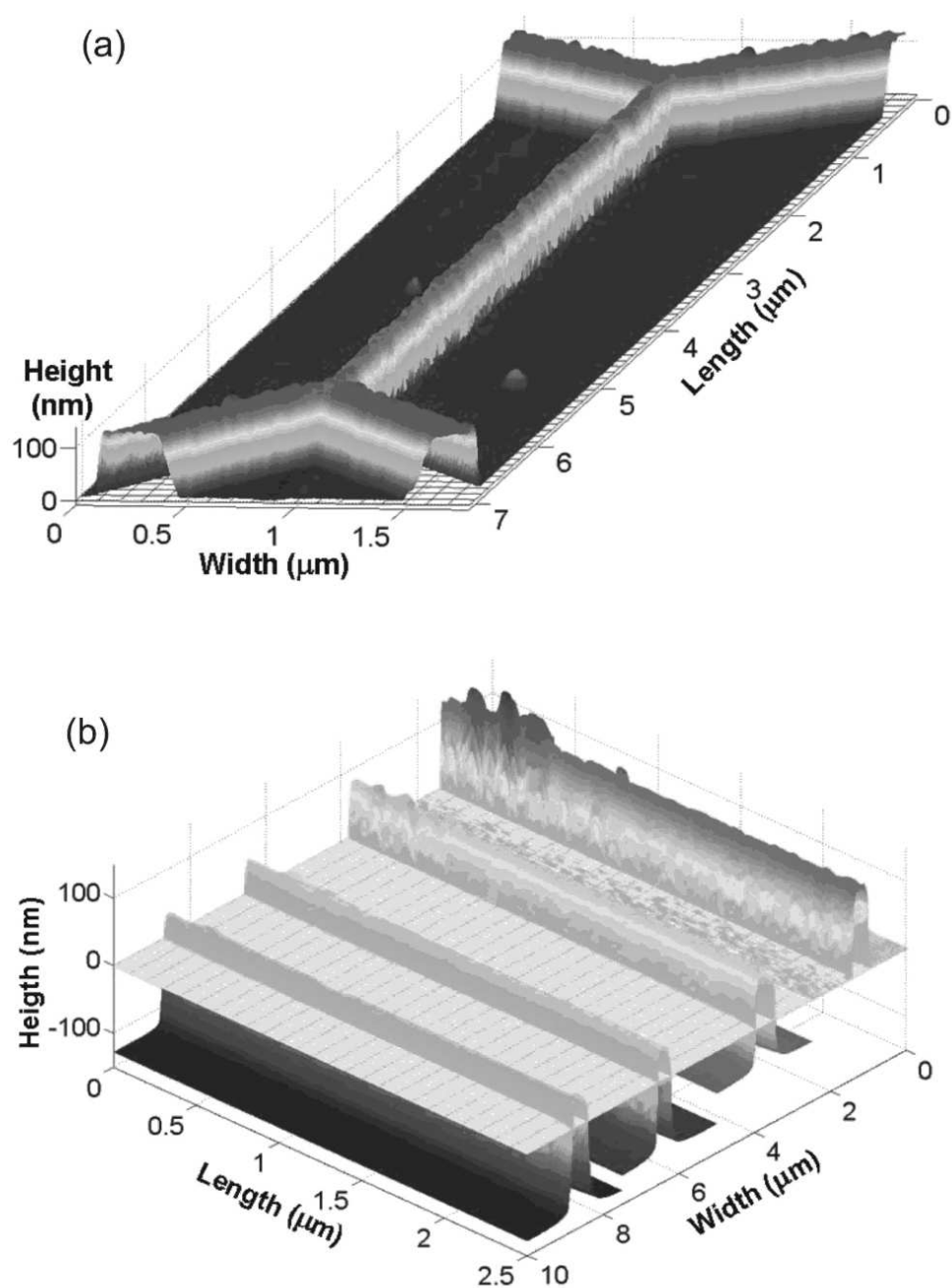}
\caption {(a) Atomic force microscope (AFM) image of a typical
5 $\mu m$ long aluminum nanowire just after lift-off. (b) AFM images showing
evolution of the shape of the same nanowire after several sessions of ion
beam sputtering. Bright color above the horizonatal plane (initial level of
the substrate) corresponds to metal, dark color below to sputtered $Si$
substrate. Note the reduction of the initial surface roughness of the nanowire.}
\end{figure}

\clearpage
\section{\protect\bigskip Results and discussion}

After a sequence of sputterings [alternated with $R(T)$ measurements] the wire
cross section $\sigma $ is reduced down to $\leq $ 20 nm, where
deviations from TAPS behavior become obvious (Fig. 2). Similar results were
obtained on several sets of aluminum nanowires with length $L$ equal to 1 
$\mu m$, 5 $\mu m$ and 10 $\mu m$. For larger diameters (e.g., Fig. 3, insert,
55 nm wire) the shape of the R(T) dependence can be qualitatively described
by the TAPS mechanism. Notice that a size dependent variation of the critical
temperature of a nanowire \cite{Shanenko}, especially pronounced in aluminum 
\cite{Zgirski NanoLett}, results in a broadening of the $R(T)$ transition and
significantly reduces straightforward applicability of the TAPS model 
\cite{Zgirski PRB limitation}. Confidence in assigning our results to the 
manifestation of a new (non-thermal) mechanism comes from the experimental
observation that in aluminum nanowires (films) the critical temperature
increases with a decrease of the wire diameter (film thickness) 
\cite{Shanenko}, \cite{Zgirski NanoLett}. It means that the broadening of the 
$R(T)$ dependencies below the bulk $T_{c}$ value $\sim $ 1.2 K cannot be
explained by a geometrical imperfection (e.g., constriction) of an aluminium nanowire. 
Below a certain limit $\sigma ^{1/2}\leq $ 20 nm fits by
the TAPS model fail to provide any reasonable quantitative agreement with
experiment  even assuming the existence of unrealistically narrow constrictions not observed by SPM (Fig. 2).

\clearpage
\begin{figure}
\includegraphics [width=1\textwidth] {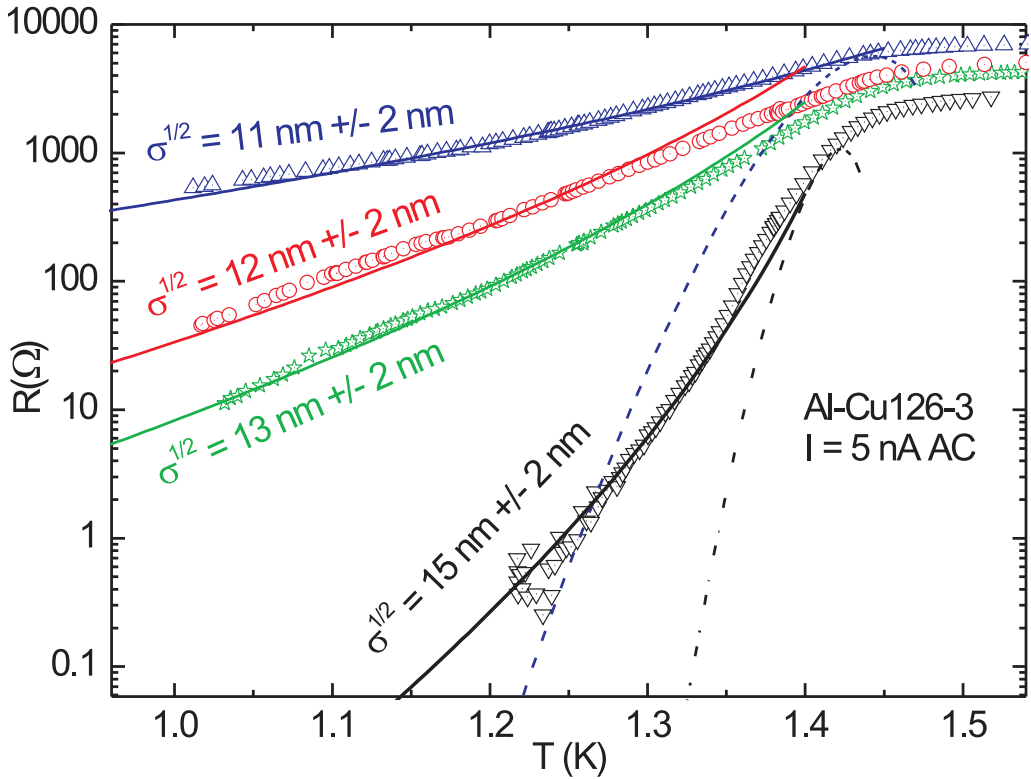}
\caption {Resistance vs temperature for the thinnest samples
obtained by progressive reduction of the diameter of the same aluminum
nanowire Al-Cu126-3 with length $L$= 10 $\mu m$. The Langer-Ambegaokar-McCumber-Halperin (LAMH) model fitting is
shown with dashed lines for 11 and 15 nm samples with the best fit mean free
path $\ell =$ 3 and 10 nm, correspondingly, $T_{c}=$ 1.46 K and critical
magnetic field $B_{c}(0)=$ 10 mT. Fitting using a simplified short wire model Ref.
\cite{Zaikin} [Eqs. 2 and 3] is shown with solid lines. For 11, 12, 13 and
15 nm wires the fitting parameters are: $T_{c}=$ 1.5 K; $A=$ 0.15, mean free
path $\ell =$ 5.4 nm, 5.8 nm, 7.3 nm, 7.5 nm; and the normal state resitance 
$R_{N}=$ 7200 k$\Omega $, 5300 k$\Omega $, 4200 k$\Omega $ and 2700 k$\Omega $.}
\end{figure}

\clearpage
Though several theoretical approaches \cite{Saito} - \cite{Pesin} were
proposed to describe the non-TAPS mechanism of finite resistance in 1D
superconducting channels, not all models are suitable for direct comparison
with experiment, lacking a clear expression for the resistance (voltage) as a
function of temperature. Where possible, we tried to compare our data with
the models and found the best agreement with renormalization theory \cite%
{Zaikin}. The full version of the model for systems of an arbitrary length $L
$ is rather sophisticated, requiring knowledge of parameters not easily
deductible from experiment. However, if the wire is short enough that only
one phase slip event can happen at a time, one can neglect the effects of the
interaction between the phase slips. In this limit the simplified expression
for the rate of QPS activation is

\begin{equation}
\Gamma _{QPS}=\frac{S_{QPS}}{\tau _{0}}\frac{L}{\xi }exp(-S_{QPS}),
\end{equation}%
where the action $S_{QPS}=A(R_{Q}/\xi )/(R_{N}/L)$, with $A$ being a
numerical constant, $R_{Q}=h/(4e^{2})=$ 6.47 $k\Omega $ and $\tau _{0}\sim
h/\Delta $ is the characteristic response time of a superconducting system
which roughly determines the duration of each QPS \cite{Zaikin}. The effective
(time averaged) voltage $V_{eff}$ due to the fluctuations can be found using the
Josephson relation \cite{Langer}. Finally, for the effective resistance of a
quasi-1D superconducting nanowire one gets

\begin{equation}
R_{QPS}(T)\equiv V_{eff}/I=\frac {h\Gamma _{QPS}}{2eI}.
\end{equation}

Contrary to TAPS \cite{Langer}, the QPS contribution has a rather weak
temperature dependence far from the critical temperature and should produce
a finite resistance even at $T\rightarrow 0$ for sufficiently narrow wires 
\cite{Zaikin}. One can achieve reasonable agreement between the model [Eqs.
2 and 3] and the experiment (Fig. 2). There are four fitting parameters:
critical temperature $T_{c}$, normal state resistance $R_{N}$, mean free
path $\ell $ [to re-calculate the dirty limit coherence length $\xi
=0.85(\xi _{0}\ell )^{1/2}$ ], and the numerical parameter $A$ of the order
of unit \cite{Zaikin}. The critical temperature and the normal state
resistance can be trivially deduced from experimental $R(T)$ dependences.
Roughly the mean free path $\ell $ can be estimated from the normal state
resistivity $\rho$, as the product $\rho \ell =5\times 10^{-16}\Omega m^{2}$
is a well-tabulated value for dirty-limit aluminum. In our 
ultra-narrow nanowires the cross-section is known from SPM measurements with 
$\pm $ 2 nm accuracy. Hence, there is some freedom in the selection of the mean free
path. As a rule of thumb for all our nanowires with effective diameter $%
\sigma ^{1/2}\leq $ 20 nm the best-fitted mean free path (at low
temperatures) was found to be roughly equal to one-half of the diameter
(Fig. 2, caption). The observation is quite reasonable, taking into
consideration that at these scales and temperatures electron scattering
is mainly determined by the sample's physical boundaries. For all aluminum
nanowires the best-fitted value for the parameter $A$ was found to be equal
to 0.15. As the value of $A$ cannot be calculated within the model \cite
{Zaikin} with the required accuracy, we believe that the correspondence between the
simplified "short wire" model and the experiment can be considered as
good.

No Coulomb blockade has been observed on $V(I)$ characteristics, indicating
good homogeneity of the wires (absence of tunnel barriers). Below the
critical temperature $V(I)$ dependences show $\sim sinh(I/I_{0})$ behavior
similar to both TAPS [Eq. 1[] and renormalization \cite{Zaikin} models. At
sufficiently small currents $I\ll 2I_{0}\sim $ 20 nA one might expect
a linear response, while at much higher currents trivial overheating is
observed. However, there exist an intermediate regime between linear and
strongly non-linear regimes: the top part of the $R(T)$ transition is not
shifted, while the slope of the bottom part decreases with an increase of the
measuring current (Fig. 3). We associate this peculiarity with the reduction
of the potential barrier $\Delta F$ by a finite current $I$ [Eq. 1], which
is qualitatively similar for thermal \cite{Langer} and quantum \cite{Zaikin}
mechanisms.
\clearpage
\begin{figure}[h]
\includegraphics[width=1\textwidth]{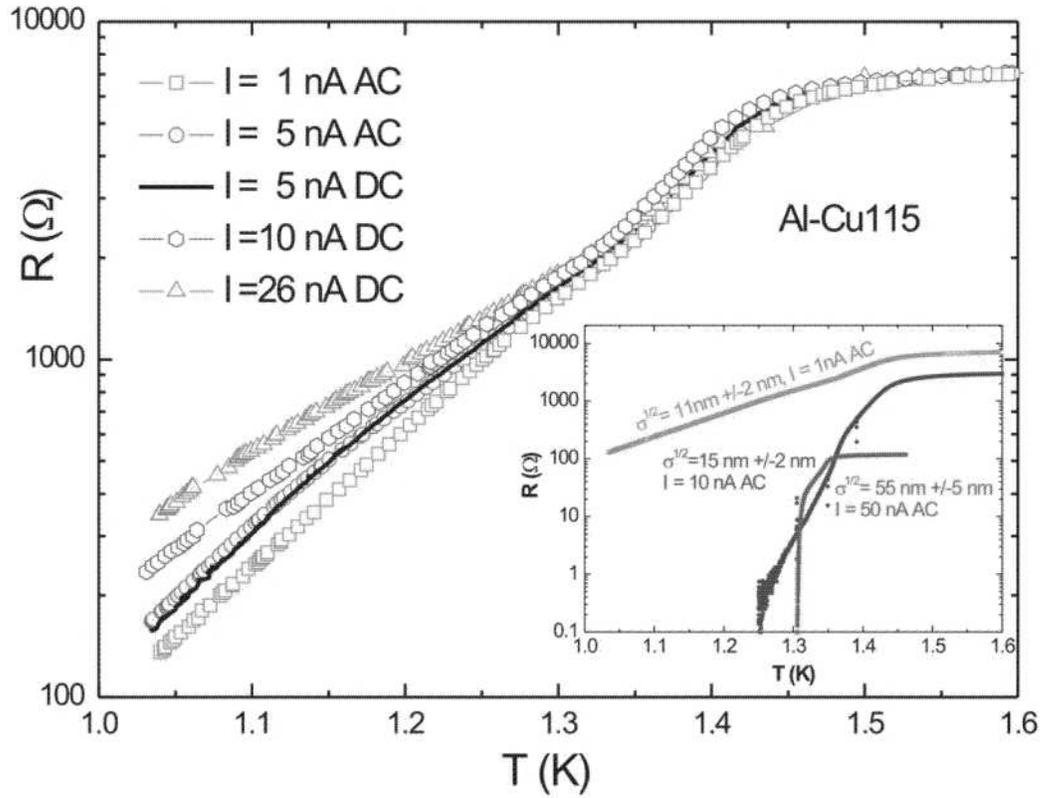}
\caption {Resistance vs temperature for a $\sigma ^{1/2}=11$ nm $%
\pm $ 2 nm sample of aluminum nanowire Al-Cu115 with length $L$= 10 $\mu m$
measured at various DC and AC currents. Inset: resistance vs temperature for
three samples obtained by progressive reduction of the diameter of the same
nanowire.}
\end{figure}
\clearpage

At certain limits the renormalization model \cite{Zaikin} predicts the same
functional dependence of the effective resistance on temperature and
current. In the high temperature limit $T\gg \phi _{0}I/k_{B}$ , $R\sim
T^{2\gamma -2}$, and in the high current limit $I\gg I_{0}$ , $R\sim
I^{2\gamma -2}$. The dimensionless conductance $\gamma =R_{Q}/R_{qp}$ is
related to the effective "quasiparticle" resistance $R_{qp}$, being
associated with dissipation provided by the quasiparticle channel. $R_{qp}$
can be considered as a fitting parameter and should be of the order of the
normal state resitance $R_{N}$. In the metallic phase, when quantum fluctuations
dominate the behavior of the system, the residual resistance at $%
T\rightarrow 0$ should correspond to $R_{qp}$ \cite{Zaikin}. The results of
fitting of our $R(T)$ and $V(I)$ dependences to the above relations are
presented in Fig. 4. One may obtain satisfactory agreement with experimental
data, allowing the quasiparticle resistance $R_{qp}$ to be smaller than the
normal state resistance $R_{N}$ of the nanowire. For each sample, the
deviation of the fitting parameter $R_{qp}$ from $R_{N}$ is less pronounced
for $V(I)$ characteristics. Probably, this discrepancy is the result of the
experimental difficulty to satisfy the applicability condition $T\gg \phi
_{0}I/k_{B}$, while the temperature should be noticeably below the critical
temperature $T_{c}$, where the whole concept of phase slippage is valid.
\clearpage
\begin{figure}[h]
\includegraphics{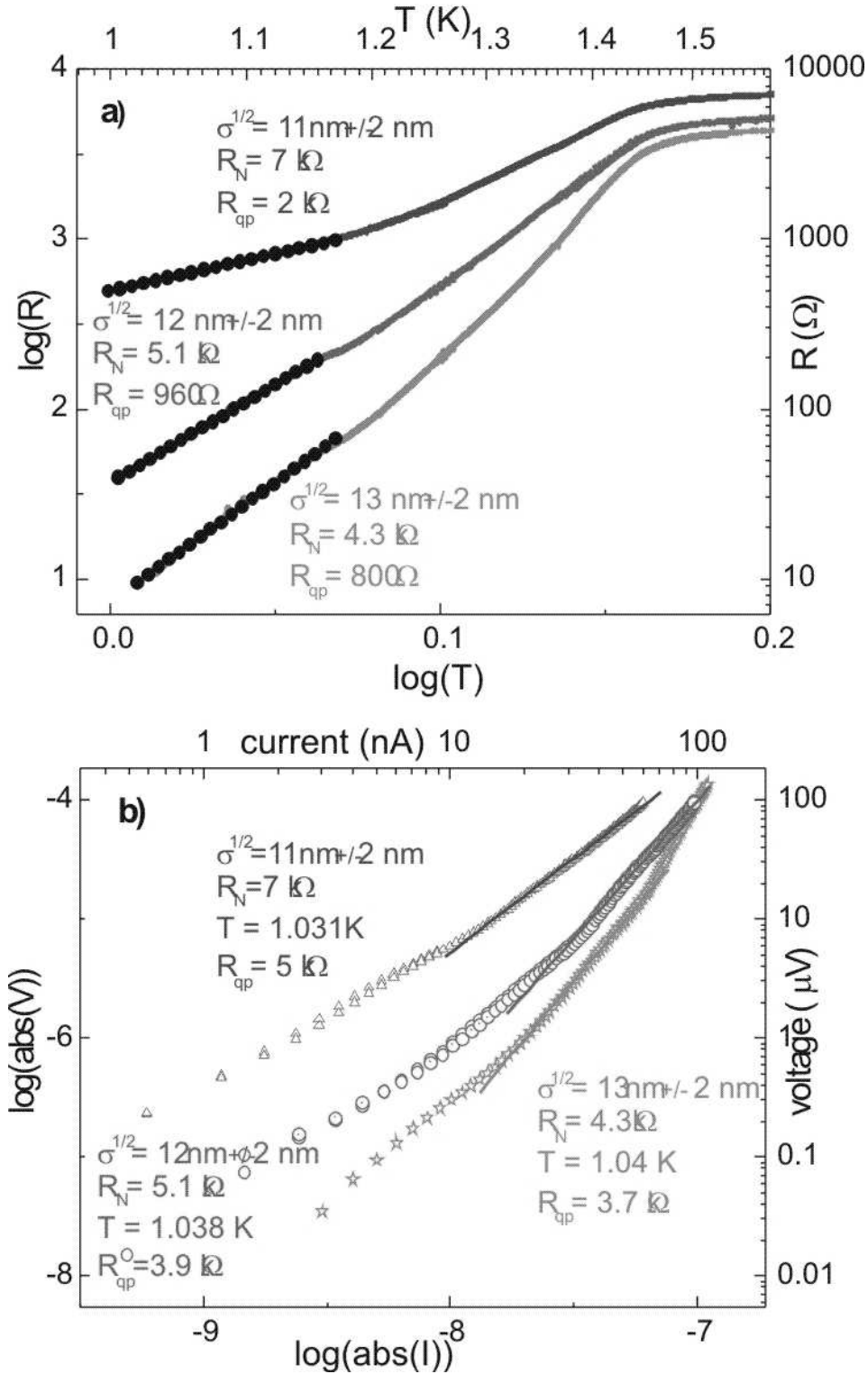}
\caption { (a) Resistance vs. temperature for three samples of the
same wire as in Fig. 2 with progressively reduced cross-sections. Solid dots
are fittings to power dependence $R\sim T^{2\gamma -2}$. (b) V(I)
dependences of the same samples taken at close temperatures stabilized with
accuracy $\pm $ 0.1 mK. Solid lines correspond to proportionality $%
R(T)\equiv V(T)/I\sim I^{2\gamma -2}$. In both figures the only fitting
parameter is the quasiparticle resistance $R_{qp}$.}
\end{figure}
\clearpage

Application of a magnetic field perpendicular to the plane of the structures
reveals a rather unusual effect. At sufficiently small temperatures a pronounced
negative magnetoresistence (nMR) is observed (Fig. 5, inset), resulting in a
"sharpening" of the $R(T)$ transitions in magnetic fields $\leq $ 25 mT
(Fig. 5). nMR is observed only in the thinest samples $\sigma ^{1/2}\leq $
20 nm and only at temperatures well below the onset of superconductivity.
A similar effect has been reported for ultrathin 1D lead strips \cite{Sharifi}. 
Aluminum is known to be immune to the creation of localized magnetic moments
provided by the majority of magnetic elements. Our samples were e-beam
evaporated from 99.999 \% pure aluminum target in an UHV chamber where magnetic
materials have never been processed. Hence, the residual concentration of
magnetic ions should be negligible. Even if we were to assume their presence,
magnetic fields where nMR is observed are too small to polarize them.
Additionally, we do not see an enhancement of $T_{c}$ related to the onset of 
superconductivity, which would be the consequence of suppression of the Kondo 
mechanism by a magnetic field. As the presence of localized magnetic moments in
our samples is not obvious, a recently developed model of nMR Ref. \cite{Bezryadin
nMR} is not applicable. nMR has been predicted in disordered superconducting
wires \cite{Pesin}. However, the main result of that model is an exponentially
small addition to the TAPS effective resistance. This conclusion contradicts 
our main observation (Fig. 2) and, hence, the applicability of the whole model 
\cite{Pesin} to our experiments is questionable. For the moment we do not
have a solid explanation for the nMR effect. The phenomenon might be related
to an interplay between two field-dependent contributions: $\Delta F$
barrier reduction and suppression of the superconducting energy gap $\Delta $.
The first process provides an expected increase of the observed resistance,
while the second one leads to a reduced quasiparticle resistance $R_{qp}$ due
to thermal activation of extra quasiparticles resulting in lower effective
resistance $R(T)$. There might be a region of magnetic fields where the
second mechanism "wins" reducing the influence of quantum fluctuations
leading to nMR \cite{Zaikin}. Another explanation of the nMR might be related to the formation of a charge imbalance region accompanying each phase slip event  \cite{Arutyunov nMR}. %%@
This non-equilibrium region provides dissipation outside the core of a phase slip. The corresponding Ohmic contribution can be effectively  suppressed by the magnetic field, resulting in %%@
nMR. However, so far the validity of the charge imbalance concept was only demonstrated  at temperatures sufficiently close to $T_c$ and its applicability to QPS at temperatures well %%@
below the critical one is not obvious. A quantitative comparison with the experiments
requires further elaboration of the theory.
\clearpage
\begin{figure}[h]
\includegraphics[width=1.1\textwidth]{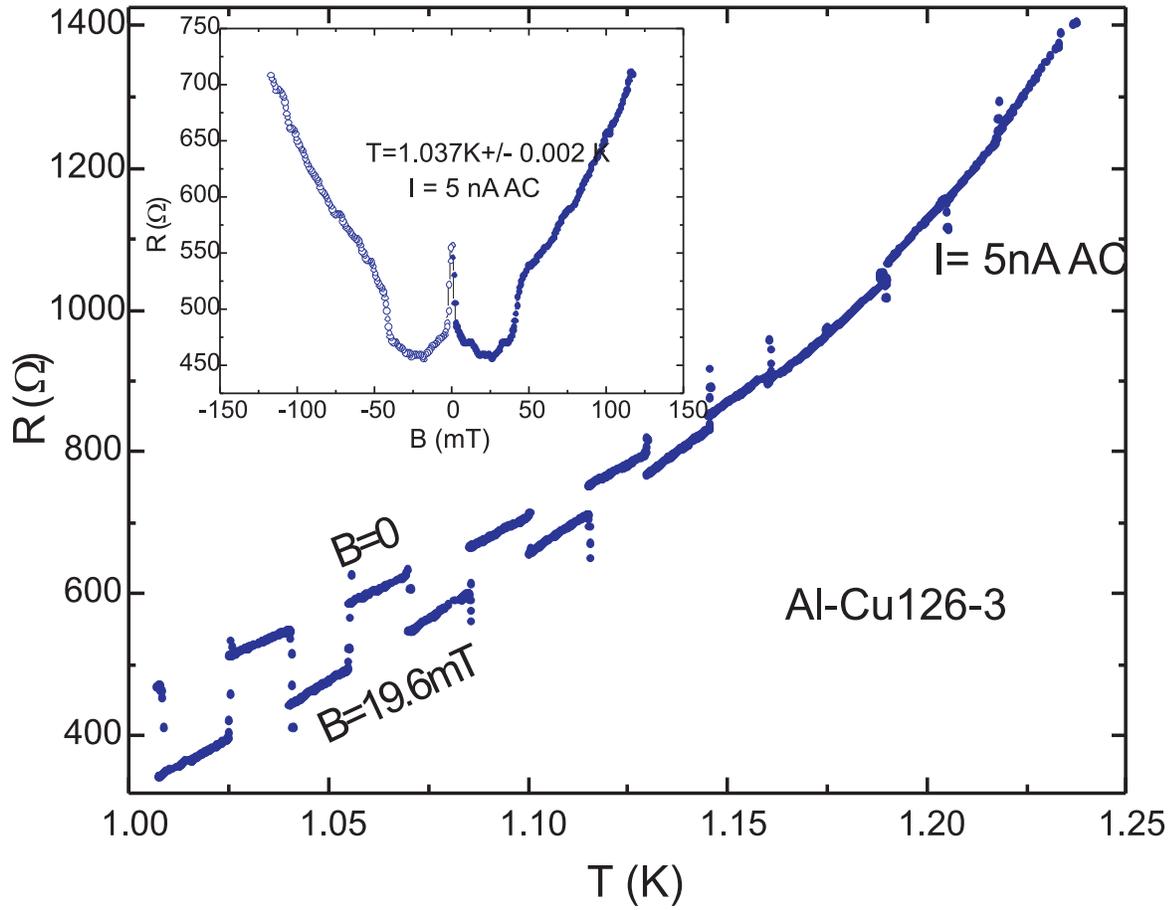}
\caption {
Slowly recorded ($\sim $ 1 h) resistance vs
temperature dependence for the 11 nm sample from Fig. 2. While sweeping the
temperature, a few times the perpendicular magnetic field $B=$ 19.6 mT was
switched on and off. The top branch corresponds to zero field, while the lower
one to field "on". Inset: resistance vs perpendicular magnetic field for
the same sample measured at constant temperature and small ac current.}
\end{figure}

\clearpage
\section{\protect\bigskip Conclusions}

The homogeneity of the wires is the central point in an interpretation of data related to phase slip mechanisms: thermal or quantum. The existence of trivial structural or geometrical %%@
imperfections as constrictions, boundaries and contact regions might broaden the $R(T)$ dependences \cite{Zgirski PRB limitation} and can be erroneously interpreted as a "new physic"'. %%@
The specially developed method of ion beam sputtering \cite {ion sputtering}, \cite{Zgirski Nanotechnology 2008} to a large extent allowed us to study the evolution of the size %%@
phenomenon, eliminating artifacts related to the uniqueness of samples fabricated in independent processing runs. The ion beam treatment polishes the surface of the samples, removing the %%@
inevitable roughness just after fabrication 
(Fig. 1). If there were no detectable geometrical imperfections in the original (thick) wires, they cannot be introduced while reducing  the diameter by the low energy ion sputtering. Formally, %%@
it cannot be excluded that in the original structures there were "hidden" undetectable structural defects (e.g., highly resistive grain boundaries) which did not contribute to conductivity, %%@
being shunted by the bulk. When the diameter of the wire is reduced below a certain scale, this type of imperfections might determine the behavior of the quasi-1D system. However, the %%@
absence of Coulomb blockade and/or gap structure on $I-V$ characteristics rules out the presence of the highly resistive tunnel barriers. 

The extensive SPM and SEM analysis indicates that after several sessions of ion beam sputtering the surface of our nanowire is flat with few nm accuracy \cite{Zgirski Nanotechnology %%@
2008}. This is quite small in absolute terms, particularly keeping in mind that the lattice constant in aluminum is $\sim$ 0.4 nm. However, approaching the 10 nm scale, a few nm roughness %%@
means a lot in relative terms. Hence, we cannot exclude the possibility that the specified average effective diameter of our nanowires is not very representative: the phase slippage might %%@
happen in the weakest (thinnest?) parts. To eliminate completely this uncertainty one should study infinitely long atomically flat structures, which are quite problematic to fabricate. %%@
Nevertheless, the main conclusion of the present paper is still valid even assuming the presence of inevitable imperfection of the samples. At all realistic parameters characterizing the %%@
geometry of the possible constrictions, the thermal activation scenario \cite{Langer} fails to explain the broad $R(T)$ dependences observed in sub-15 nm nanowires, while the QPS %%@
mechanism \cite {Zaikin} provides good agreement with the experiment (Fig. 2).

In conclusion, we have traced the evolution of the shape of the superconducting
transition $R(T)$ in quasi-one-dimensional aluminum nanowires with
progressive reduction of their cross-sections. For relatively thick samples $
\sigma ^{1/2}\geq $ 20 nm the shape of the transitions can be qualitatively
explained in terms of wire inhomogeneity and the model of thermally
activated phase slips (TAPS). While in nanowires with effective diameter 
$\leq $ 15 nm the $R(T)$ dependences are much wider and no reasonable set
of fitting parameters can account for the TAPS mechanism. The phenomenon is
associated with manifestation of quantum fluctuations of the order
parameter. The results are in good agreement with renormalization
theory \cite{Zaikin}. The effect of quantum fluctuations should have a
universal validity, indicating a breakdown of the zero resistance state in
quasi-one-dimensional superconductors, setting fundamental limitations on
miniaturization of nanoelectronic components designed to carry a
dissipationless supercurrent.

\section{\protect\bigskip Acknowledgments}
The authors would like to acknowledge C. N. Lau, A. Bezryadin, D. Golubev,
L. Pryadko, A. Zaikin, and D. Vodolazov for their helpful discussions. The
work was supported by the EU Commission FP6 NMP-3 project 505457-1 ULTRA-1D
"Experimental and theoretical investigation of electron transport in
ultranarrow 1-dimensional nanostructures".

%%%%%%%%%%% FIGURES %%%%%%%%%%%%%%%%%%%%%%%%%%%%

%%%%%%%%%%% BIBLIOGRAPHY %%%%%%%%%%%%%%%%%%%%%%%

\end{document}